\documentclass[aps,pre,twocolumn,nofootinbib,floatfix,showpacs]{revtex4}  

\usepackage{color}
\usepackage{soul}

\usepackage{graphicx} 
\usepackage{amssymb}
\newcommand {\be}{\begin{equation}}
\newcommand {\ee}{\end{equation}}
\newcommand {\ba}{\begin{eqnarray}}
\newcommand {\ea}{\end{eqnarray}}

\newcommand{\Jm}{J_{\mathrm{m}}}
\newcommand{\Om}{\omega_{\mathrm{m}}}
\newcommand{\Os}{\omega_{\mathrm{l}}}
\newcommand{\kl}{k_{_L}}
\newcommand{\kr}{k_{_R}}
\newcommand{\kc}{k_{_C}}
\newcommand{\ko}{k_{_0}}
\newcommand{\Vl}{V_{_L}}
\newcommand{\Vr}{V_{_R}}
\newcommand{\nc}{n_{_c}}
\newcommand{\Jl}{J_{_L}}
\newcommand{\Jr}{J_{_R}}

\begin{document}


\title{Energy transport in harmonically driven segmented Frenkel-Kontorova lattices}

\author{M.~Romero-Bastida}
\affiliation{SEPI ESIME-Culhuac\'an, Instituto Polit\'ecnico Nacional, Av. Santa Ana No. 1000, Col. San Francisco Culhuac\'an, Delegaci\'on Coyoacan, Distrito Federal 04430, Mexico}
\email{mromerob@ipn.mx}

\date{\today}

\begin{abstract}
In this work we study the energy transport in a one-dimensional system composed of two dissimilar Frenkel-Kontorova lattices connected by a time-modulated coupling and in contact with two heat reservoirs operating at different temperature by means of molecular dynamics simulations. There is a value of the driving frequency at which the heat flux takes its maximum value, a phenomenon termed as thermal resonance. Structural modifications in the lattice strongly alter the way in which the external driving interacts with the phonon bands. The overlap of the latter in the harmonic regime of the model determine the frequency range wherein resonance emerges. Parameter dependencies by which the incoming heat flux can be directed to either of the heat reservoirs are examined as well. Our results may be conductive to further developments in designing thermal devices. 
\end{abstract}



\maketitle

\section{Introduction\label{sec:Intro}}

Structural stability in mesoscale and nanoscale devices is closely related to heat generation within. Experiments measuring the heat generated in electric current-carrying metal-molecule junctions determined that the generated heat can be substantial~\cite{Tsutsui08,Huang07,Tsutsui19} and can therefore threaten the device's integrity. Efficiently dissipating heat in such devices is thus important and a problem that must be considered, especially with regard to technological applications. An example of devices wherein the above issues are relevant is that of microelectromechanical and nanoelectromechanical systems. These are being developed for a host of nanotechnological applications, such as highly sensitive mass~\cite{Yang06}, spin~\cite{Rugar04}, and charge detectors~\cite{Andrew98}, as well as for basic research in the mesoscopic physics of phonons~\cite{Volz16}. In fact, an improved understanding of the manipulation and control of phonons ---that manifest themselves as heat at the nanoscale level--- is necessary for further progress in addressing the above mentioned problem~\cite{Li12}.

Several models and mechanisms have been proposed to control or manipulate the heat flux at the nanoscale. For a static thermal bias the most explored control mechanism so far has been by means of tuning the structural asymmetry and the degree of anharmonicity in tailored one-dimensional lattice structures. The ensuing dependence on temperature of the power spectra of dissimilar segments results in the phenomenon of thermal rectification, i.e. asymmetrical heat flow, with substantial progress being achieved in the last two decades~\cite{Roberts11}. This phenomenon offers improved thermal management at the nanoscale and its success can be gauged by the fact that, only shortly after the first theoretical models of the heat rectifier~\cite{Terraneo02} and thermal memory~\cite{Wang08} were developed, successful experimental realizations were reported~\cite{Chang06a,Xie11}.

In order to obtain an even more flexible control of heat energy, comparable with the richness available for electronics, one may utilize temporal modulation that directs heat from one part of the device to another or to an external reservoir by means of an applied external work. Models on the mechanism of such a nanoscale heat pump have been proposed in systems, mostly coupled anharmonic lattices, where there is no net thermal bias between the two reservoirs~\cite{Zhan09,Nianbei09,Ren10} and where the pump works against the imposed static thermal gradient~\cite{Broek06,Bao-quan10,Zhang11}. Furthermore, models based on pumping phenomena have been proposed as moving barriers in a cavity to pump phonons from a cold reservoir to a hotter one, or driven two-level systems or molecular junctions in asymmetric contact with phononic baths characterized by different spectral properties~\cite{Arrachea2012,Segal06}. In addition, other models employing quantum particle pumps that differentiate and filter hot and cold particles have been proposed~\cite{Rey07}. From the above examples it can be inferred that among the necessary prerequisites to run such heat machinery are nonlinearity, thermal noise, unbiased nonequilibrium driving, and a symmetry-breaking mechanism. However, other than the aforementioned information, externally driven energy transport remains poorly understood. Furthermore, since its study is far from trivial, contradictory results have been reported. For example, in Ref.~\cite{Bao-quan10} it was claimed that heat pumping appeared in the Frenkel-Kontorova (FK) lattice under the influence of a periodic driving force, but later it was demonstrated that such an effect is indeed absent~\cite{Zhang11}.

In this work we reconsider the one-dimensional model consisting of two dissimilar FK lattices connected together by a time-modulated harmonic coupling under the influence of a static thermal bias previously studied in Ref.~\cite{Beraha15}. Our results seem to indicate that the interpretation of some of the results in the aforementioned work may be questionable. More precisely, the obtained resonant heat transport regime, i.e. maximization of heat flux for a specific value of the external driving frequency, was explained by a shift towards lower frequencies of the phonon bands in each segment, induced by a transition from anharmonic to harmonic behavior as the resonant frequency is approached. Nevertheless, we have observed that such a phonon band shift can only be attained by the imposed structural asymmetries of the model. We have introduced them in a systematic way by means of a scaling of the involved parameters, first proposed in Ref.~\cite{Li04a} to study the thermal rectification properties of this model and extensively employed afterwards~\cite{Hu06,Li12}. Once the asymmetry is fixed, no shift in the phonon bands is observed altogether; these are only weakly modified by the external drive. Furthermore, we have determined that, for the temperature bias employed in Ref.~\cite{Beraha15} and in the present work, the system is in its {\it harmonic} regime. The overlap of the ensuing phonon bands determines the frequency range wherein thermal resonance can manifest itself, in close analogy to the way such overlap is at the origin of thermal rectification in this and similar systems. Thus we unravel the underlying physical mechanism for such a resonance phenomenon.

This paper is organized as follows: in Sec.~\ref{sec:Model} the model system and methodology are presented. Our results on the dependence of the thermal resonance on the structural parameters of the model are reported in Sec.~\ref{sec:Res}. The discussion of the results, as well as our conclusions, are presented in Sec.~\ref{sec:Disc}.

\section{The Model\label{sec:Model}}

\begin{figure}\centering
\includegraphics[width=0.85\linewidth,angle=0.0]{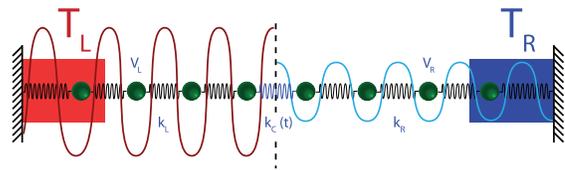}
\caption{(Color online) Sketch of our model system composed of two dissimilar FK lattices connected by a time-modulated harmonic interaction and in contact with two thermal reservoirs.}
\label{fig:1}
\end{figure}

Our system consists of two segments ($L,R$) of nonlinear 1D lattices coupled together by a harmonic spring with a time-modulated strength $\kc(t)$, as shown in Fig.~\ref{fig:1}. The equations of motion (EOM) for a given oscillator within each segment can be written, in term of dimensionless variables, as $\dot q_i =p_i/m_i$ and
\ba
\dot p_i&=&k_{_{L,R}}(q_{i+1} + q_{i-1} - 2q_i) - {V_{_{L,R}}\over2\pi}\sin(2\pi q_i) \cr
   & + & (\xi_{_1} - \gamma_{_1} p_i)\,\delta_{i1} + (\xi_{_N} - \gamma_{_N} p_i)\,\delta_{i{N}},
\ea
where $N$ is the system size. $k_{_{L,R}}$ and $V_{_{L,R}}$ are the harmonic spring constant and the amplitude of the FK on-site potential in each segment, respectively; more precisely, in the above equations we employ $\kl,V_{_{L}}$ if $i\in[1,\nc-1]$ and $\kr,V_{_{R}}$ if $i\in[\nc+2,N]$. In order to reduce the number of adjustable parameters, we set $\Vr=\lambda\Vl$ and $\kr=\lambda\kl$. Here we consider only the commensurate case where the on-site potential assumes the same spatial periodicity as the lattice constant. $\{m_i,q_i,p_i\}_{i=1}^{N}$ are the dimensionless mass, displacement, and momentum of the $i$th oscillator; see the Appendix of Ref.~\cite{Li12} for a detailed procedure on how to construct such dimensionless variables. Fixed boundary conditions are assumed ($q_{_0}=q_{_{{N}+1}}=0$). Henceforth we will consider a homogeneous system, i.e., $m_i=1\,\,\forall\,i$. The Gaussian white noise $\xi_{_{1,N}}$ has zero mean and correlation $\langle\xi_{_{1,N}}(t)\xi_{_{1,N}}(t^{\prime})\rangle=2\gamma_{_{1,N}}k_{_B}T_{_{1,N}}m_i(\delta_{1i}+\delta_{{N}i})\delta(t-t^{\prime})$, with $\gamma_{_{1,N}}$ (taken as $0.5$ in all computations hereafter reported) being the coupling strength between the system and the left and right thermal reservoirs operating at temperatures $T_{_L}=0.15$ and $T_{_R}=0.05$, respectively; the system thus operates at a constant average temperature value of $T_{_0}\equiv(T_{_L}+T_{_R})/2=0.1$. The EOM for the last oscillator in the first segment ($L$) and the first one in the second ($R$) are given by
\ba
\dot p_{\nc}&=&\kl(q_{{\nc}-1} - q_{\nc}) + \kc(t)(q_{{\nc}+1} - q_{\nc}) \cr
     & - & {\Vl\over2\pi}\sin(2\pi q_{\nc}), \cr
\dot p_{{\nc}+1}&=&\kr(q_{{\nc}+2} - q_{{\nc}+1}) + \kc(t)(q_{\nc} - q_{{\nc}+1}) \cr
     & - & {\Vr\over2\pi}\sin(2\pi q_{{\nc}+1}),
\ea 
with $\kc(t)=\ko (1+\sin\omega t)$ being the time-modulated amplitude of the harmonic coupling, which is an external driving with frequency $\omega$. The aforementioned EOM were integrated with a stochastic velocity-Verlet integrator with a time step of $\Delta t=0.005$ for a production time interval of $2\times10^7$ time units after a transient time of $10^{8}$ time units.

Once the nonequilibrium stationary state is attained, the local heat flux is computed as ${J}_i=k_{_{L,R}}\langle\dot q_{i}(q_{i}-q_{i-1})\rangle$, with $\kl$ if $i\in[2,\nc-1]$ and $\kr$ if $i\in[\nc+1,N]$, and the local temperature as $T_i=\langle p_i^2/m_i\rangle$; in both instances $\langle\cdots\rangle$ indicates time average. In the stationary state the heat flux in each segment becomes independent of the site, and, in order to improve the statistical precision of our results, the mean heat flux $J_{_{L,R}}$ on each side of the lattice is calculated as the algebraic average of ${J}_i$ over the number of unthermostatted oscillators in each segment. Now, the rate of work $\dot W$ done by the external driving in the contact at $\nc$ is dissipated into the reservoirs, implying that
\ba
\dot W = \Jl + \Jr,
\ea
where $J_{_{L,R}}$ are defined as positive when the heat flows into the reservoirs.

\section{Results\label{sec:Res}}

\begin{figure}\centering
\includegraphics[width=0.85\linewidth,angle=0.0]{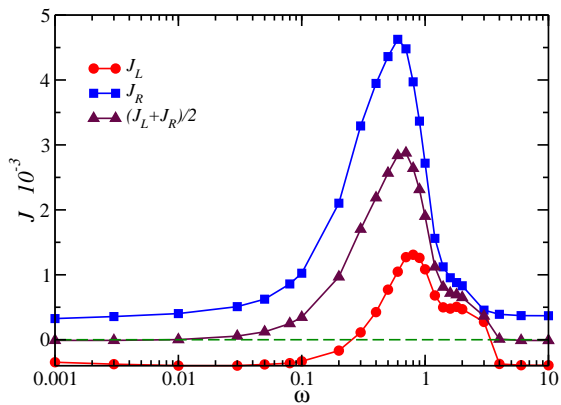}
\caption{(Color online) Heat flux vs driving frequency $\omega$. The energy currents through the left and right segments are $\Jl$ and  $\Jr$, respectively; $J\equiv(J_{_L}+J_{_R})/2$ is the average current with $\Vl=5$, $\kl=1$, $\lambda=0.2$, $\ko=0.05$, $N=32$, and $\nc=N/2$. Maximum heat flux is at $\Om=0.6$. Continuous lines are a guide to the eye.}
\label{fig:2}
\end{figure}

In Fig.~\ref{fig:2} we present the results of the dependence of heat fluxes $\Jl$, $\Jr$, and the average $J$ as a function of the driving frequency $\omega$ with $\Vl=5$, $\Vr=1$ $\kl=1$, $\kr=0.2$, $\ko=0.05$, and $\nc=N/2$ for a lattice with $N=32$ oscillators. Our results in Fig.~\ref{fig:2}(a) are qualitatively similar to those reported in Ref.~\cite{Beraha15}. The observed differences are clearly a result of the way in which we chose to implement the structural asymmetry into the system, which is that of Ref.~\cite{Li04a} as previously mentioned. In the adiabatic driving limit $\omega\rightarrow0$ the heat flows from oscillator $i=1$ to $i=N$, with $\Jr=-\Jl>0$, and thus the averaged net power released to the system is zero. In the opposite limit $\omega\rightarrow\infty$ the coupling oscillates very fast and converges to a time average constant value $\ko$ as if there is no driving. Thus the relevant phenomenology occurs at intermediate $\omega$ values. First, for $0.1<\omega\lesssim0.26$ there is a small net power contribution released to the system, and now $\Jr-|\Jl|=P>0$; thus, although heat still flows from the hot to the cold reservoir, $\Jl$ and $\Jr$ have different magnitudes. Then, within $0.26\lesssim\omega\lesssim2$, the power released in the contact region is dissipated into the reservoirs, since $\Jl>0$. This phenomenon, due to the resonant interaction of the external drive with the system's intrinsic frequencies, is known as {\it thermal resonance}~\cite{Bao-quan10,Zhang11,Beraha15}. Its main feature is the maximization of the heat flux $\Jm$ at a specific frequency, $\Om=0.6$ in this particular instance.

\begin{figure}
\centerline{\includegraphics*[width=75mm]{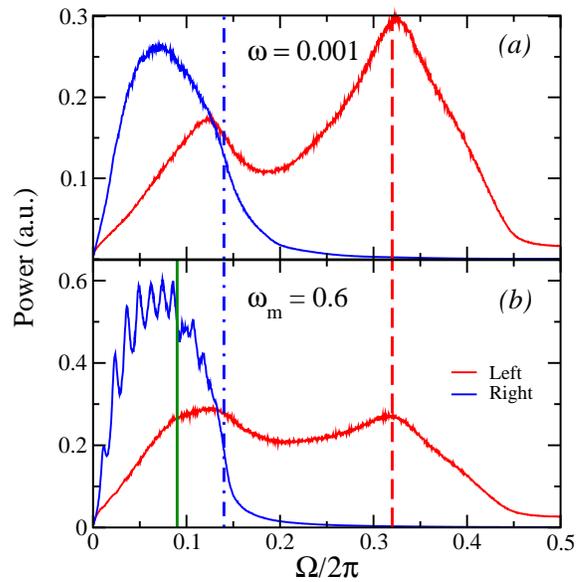}}
\caption{(Color online) (a) Power spectra of the two oscillators in the left (red) and right (blue) sides of the contact for a driving frequency of $\omega=0.001$. (b) Same as (b) but for $\Om=0.6$. Vertical dashed and dot-dashed lines correspond to the cut-off frequencies of the left and right phonon bands respectively. Vertical solid line denotes the $\Om/2\pi$ value. Same $\Vl$, $\kl$, $\lambda$, $\ko$, $N$, and $\nc$ values as in Fig.~\ref{fig:2}.}
\label{fig:3}
\end{figure}

Next we will explore the relative contribution of low- and high-frequency phonons to the heat released into the system through the contact region. In Fig.~\ref{fig:3} we plot the phonon spectra $|\tau^{-1}\!\!\int_{_0}^{\tau}\!\! dt\dot q_i(t)\exp(-\mathrm{i}\Omega t)|^2$ of the interface oscillators at the left and right sides of the contact for both the adiabatic driving limit, $\omega=0.001$, and in the thermal resonance regime, $\Om=0.6$, which corresponds to the maximum heat flux $\Jm$ of Fig.~\ref{fig:2}. It is evident that the spectra in the latter regime have almost twice the spectral power as those in the former, which results in a larger overlap in the low-frequency range of the thermal resonance regime depicted in Fig.~\ref{fig:3}(b). The energy transport into the reservoirs goes through the phonon channels determined by the imposed thermal bias since the structure, unlike the magnitude, of the spectra remains largely unchanged. The results in our Fig.~\ref{fig:3} can be understood if we recall that there is a critical temperature $T_{_{\mathrm{cr}}}\approx V/(2\pi)^2$ above which the kinetic energy is large enough to overcome the on-site potential barrier, hence the contribution of the on-site potential can be neglected.~\cite{Li04a}. In our particular case we have $T_{_{\mathrm{cr}}}^{(L)}=0.13<T_{_L}$ for $V_{_L}=5$ and $T_{_{\mathrm{cr}}}^{(R)}=0.025<T_{_R}$ for $V_{_R}=1$. Thus both sides of the system are in a temperature regime, well above their respective $T_{_{\mathrm{cr}}}$ values, wherein they behave as harmonic lattices with a phonon band of $0<\Omega<\sqrt{4k_{_{L,R}}}$ composed mainly of noninteracting phonons which gives $0<\Omega/2\pi\lesssim0.32$ for the left oscillator and $0<\Omega/2\pi\lesssim0.14$ for the right one~\cite{Li04a}. In the weak-coupling limit ($\ko\ll1$) thermal resonance can occur only for frequency values in the overlapping region of these phonon bands, which implies that $\Om/2\pi<0.14$; our result $\Om/2\pi\sim0.09$ is in good agreement with the above estimate. Therefore, the net energy flow from the external source into the thermal reservoirs is accomplished by the external driving due to its interaction with ---and ensuing alteration of--- the phonon bands activated by the thermal bias imposed at the boundaries.

\begin{figure}\centering
\includegraphics[width=0.85\linewidth,angle=0.0]{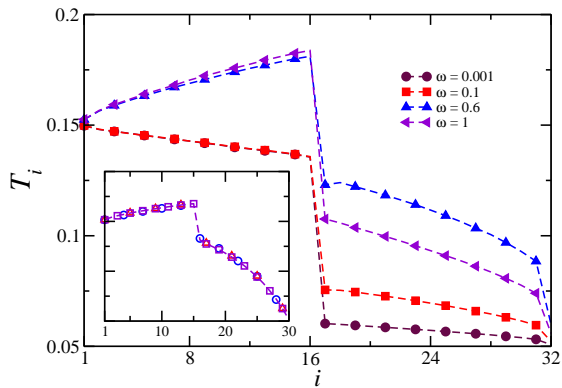}
\caption{(Color online) Temperature profiles for selected $\omega$ values, from low to high ones, increasing from bottom to top. Same $\Vl$, $\kl$, $\lambda$, $\ko$, $N$, and $\nc$ values as in Fig.~\ref{fig:2}. Inset is the temperature profile for the same values of the parameter set as in Ref.~\cite{Beraha15} with a resonant frequency of $\Om=0.3$ for production time intervals of $5\times10^6$ (circle), $2\times10^7$ (triangle), and $10^8$ (square) time units.}
\label{fig:4}
\end{figure}

In Fig.~\ref{fig:4} we plot the corresponding temperature profiles for selected $\omega$ values reported in Fig.~\ref{fig:2}. It is clear that, near and at $\Om$, there is a change in the sign of the slope corresponding to the left side of the system, which signals that there is an energy flux into the hot reservoir. The slope of the temperature profile in the right side is also increased with respect to its value in the adiabatic regime because of the additional energy flux afforded by the external driving. Due to the temperature jump at the right boundary the slope of the temperature profile is higher than that in the left side, which correlates well with the fact that $\Jr>\Jl$ for $\omega=\Om$. Now, contrary to Ref.~\cite{Beraha15}, all of the temperature profiles reported in Fig.~\ref{fig:4} exhibit a discontinuity at the interface. This result is consistent with the existence of the contact between both lattices, as shown in numerous studies~\cite{Li04a,Hu05,Luo07}. We corroborated that this temperature jump is still present when simulations were performed with the exact values of the employed parameter set reported in the aforementioned reference and different production time intervals, as can be readily seen in the Fig.~\ref{fig:4} inset.

\begin{figure}\centering
\includegraphics[width=0.85\linewidth,angle=0.0]{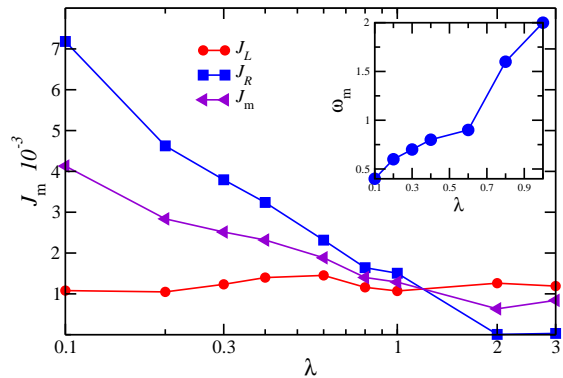}
\caption{(Color online) Heat flux $\Jm$ vs scaling parameter $\lambda$. Same $\Vl$, $\kl$, $\ko$, $N$, and $\nc$ values as in Fig.~\ref{fig:2}. Inset is $\Om$ vs $\lambda$. Continuous lines are a guide to the eye.}
\label{fig:5}
\end{figure}

As already explained in Sec.~\ref{sec:Model}, both the magnitude of the elastic constant and the strength of the on-site potential can be simultaneously controlled by means of the $\lambda$ parameter. Therefore, it is sufficient to study the properties of the considered system as a function of the aforementioned parameter. Figure~\ref{fig:5} reports the dependence of $\Jm$ versus $\lambda$. This figure clearly shows that, for $\lambda<1$, we have $\Jr>\Jl$, i.e., more heat flows into the cold reservoir. Next, as $\lambda$ increases, and thus the asymmetry in the system decreases, $\Jr$ decreases. For $\lambda>1$ values there is an almost complete suppression of the heat flux in the right side of the lattice. In this situation the hight of the valley and the harmonic constant stiffness become large enough to confine the oscillators in the potential valley, thus preventing any significant heat flux altogether into the right segment; only a steady heat flow into the left (hot) reservoir through the corresponding segment is present. In the inset we plot $\Om$ versus $\lambda$, and it can be readily observed that, as $\lambda$ increases, the resonant frequency has a sharp increase in its value when $\lambda\sim0.6$.

\begin{figure}\centering
\includegraphics[width=0.85\linewidth,angle=0.0]{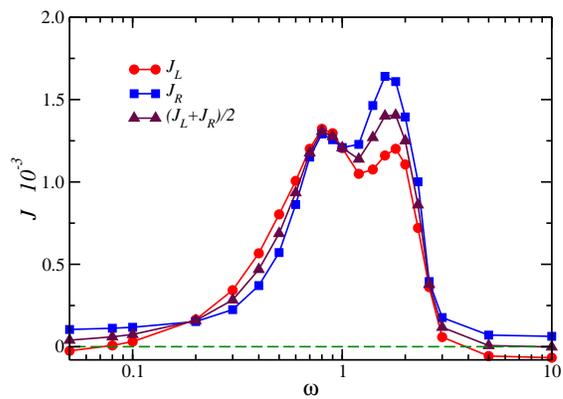}
\caption{(Color online) (a) Heat flux vs driving frequency $\omega$ for $\lambda=0.8$. Same $\Vl$, $\kl$, $\ko$, $N$, and $\nc$ as in Fig.~\ref{fig:2}. The two peaks of the heat flux correspond to frequencies $\Os=0.8$ and $\Om=1.6$. Continuous lines are a guide to the eye.}
\label{fig:6}
\end{figure}

For the instances in Fig.~\ref{fig:5} wherein $\lambda<0.6$ we have $\Om<1$, as can be seen in the inset, with a dependence of the heat flux on the driving frequency very similar to that already depicted in Fig.~\ref{fig:2}. In Fig.~\ref{fig:6} we display the data of the heat flux as a function of the driving frequency $\omega$ for the particular $\lambda=0.8$ value. It is to be noted that a peak associated with a $\omega<1$ frequency is still present at $\Os\sim0.8$, but is now accompanied by a second peak, that now becomes the absolute maximum and thus the resonant frequency, at $\Om\sim1.6$, as can be readily noticed, with $\Jr>\Jl$.

\begin{figure}
\centerline{\includegraphics*[width=75mm]{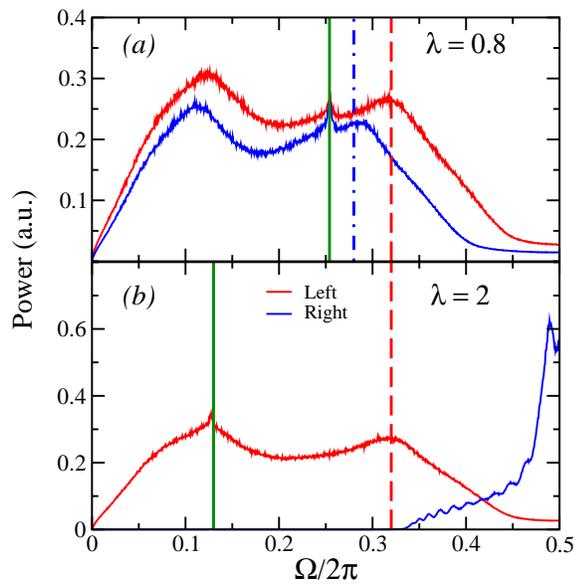}}
\caption{(Color online) (a) Power spectra of the two oscillators in the left (red) and right (blue) sides of the contact for a $\lambda=0.8$ value. (b) Same as (b), but for $\lambda=2$. Vertical dashed and dot-dashed lines correspond to the cutoff frequencies of the left and right phonon bands, respectively. Vertical solid line denotes the corresponding $\Om/2\pi$ values in each case. Same $\Vl$, $\kl$, $\ko$, $N$, and $\nc$ values as in Fig.~\ref{fig:2}.}
\label{fig:7}
\end{figure}

The power spectra of the oscillators to the left and right of the contact for the $\lambda=0.8$ value of Fig.~\ref{fig:5} are displayed in Fig.~\ref{fig:7}(a). The increased structural symmetry of the system is reflected in the qualitative similarity of the two spectra, and the ensuing overlap over the entire frequency range accounts for the widening of the range wherein subresonant behavior can be observed. The left peak of the $J$ vs $\omega$ plot in Fig.~\ref{fig:6} can be associated with the leftmost ones of the displayed spectra since they are located at $\Omega/2\pi\sim0.13$, a value almost identical to $\Os/2\pi$. However, the value for which thermal resonance appears, $\Om/2\pi\sim0.254$, coincides with a spike-like value in both spectra, as can be readily appreciated. Just as in the instance depicted in Fig.~\ref{fig:3}, $\Om$ lies in the frequency range wherein the phonon bands overlap. These are given by $0<\Omega/2\pi\lesssim0.32$ for the left side and $0<\Omega/2\pi\lesssim0.28$ for the right one; thus $\Om/2\pi<0.28$, as is indeed observed. The $\lambda=2$ instance is displayed in Fig.~\ref{fig:7}(b). As previously mentioned, now $\Vr$ becomes relevant, and thus so does the influence of the anharmonic FK potential. In this case the lower bound of the phonon band is raised by $\sqrt\Vr$ and the phonon band is shifted to $\sqrt{\Vr}<\Omega<\sqrt{\Vr+4\kr}$~\cite{Li04a}. For the considered conditions the latter is $0.5<\Omega/2\pi<0.68$, which has no possible overlap with the left phonon band $0<\Omega/2\pi\lesssim0.32$. Therefore it is clear that the only open channel available for heat carrying phonons in the low-frequency region is now afforded by the latter phonon band. And indeed the resonant frequency $\Om/2\pi\sim0.127$ lies within this last frequency range. Thus the heat flux into the left (hot) reservoir is dominated by low-frequency acoustic phonons. Finally, $\Om$ coincides approximately with the location of the highest (leftmost) peak of the spectrum and exactly with a small spike-like perturbation at that same frequency value.

\begin{figure}\centering
\includegraphics[width=0.85\linewidth,angle=0.0]{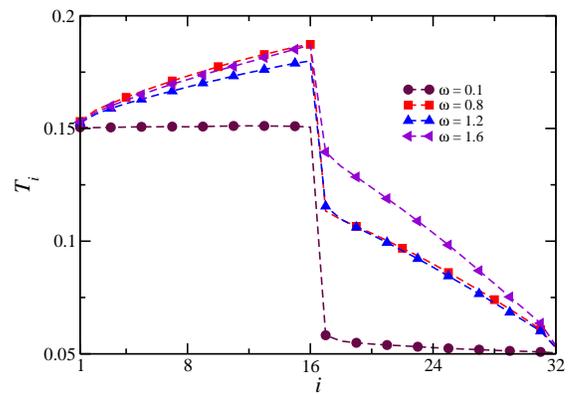}
\caption{(Color online) Temperature profiles for selected $\omega$ values, from low to high ones, increasing from bottom to top in the case of $\lambda=0.8$. Same $\Vl$, $\kl$, $\ko$, $N$, and $\nc$ values as in Fig.~\ref{fig:2}.}
\label{fig:8}
\end{figure}

The temperature profiles for various driving frequency values corresponding to the $\lambda=0.8$ case depicted in Fig.~\ref{fig:6} are plotted in Fig.~\ref{fig:8}, and an asymmetry between the two sides of the system is evident, as expected. Now for the $\omega$ values considered the slopes in the left side are very similar, indicating that the magnitude of the heat flux towards the left reservoir is almost the same in all cases, including $\omega=0.8$, i.e. the left most peak. However, in the right side the slope corresponding to $\omega=1.6$, i.e., the rightmost peak and absolute maximum, is clearly greater than the other ones, indicating a heat flux towards the right (colder) reservoir of larger magnitude than that in the left side. We also notice that the jump in the part of the temperature profile in contact with the cold reservoir that was observed for the corresponding $\Om$ case in Fig.~\ref{fig:4} is now absent. This effect is most certainly a consequence of the increased symmetry of both segments, since it is well known that the temperature profile of the homogeneous FK lattice has no discontinuities in the boundaries~\cite{Hu98}.

\begin{figure}\centering
\includegraphics[width=0.85\linewidth,angle=0.0]{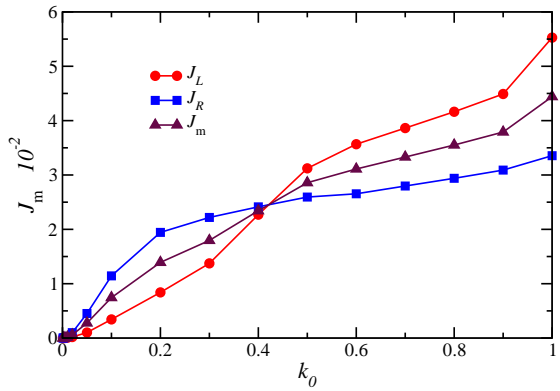}
\caption{(Color online) Heat flux $\Jm$ vs $\ko$. Same $\Vl$, $\kl$, $\lambda$, $N$, and $\nc$ values as in Fig.~\ref{fig:2}. Continuous lines are a guide to the eye.}
\label{fig:9}
\end{figure}

The interface elastic constant $\ko$ is a very important parameter as it plays the role of coupling the two lattices. By adjusting this parameter one can control the heat flow through the system, as previous work has shown~\cite{Hu06}. Indeed, once its value is fixed, then the smaller the coupling is, the smaller the heat current is through the system. In Fig.~\ref{fig:9} we present the variation of $\Jm$ as a function of $\ko$. It is clear that the average value of $\Jm$ presents a monotonic increase as the magnitude of the elastic constant grows. However, for $\ko<0.4$ values we have $\Jl<\Jr$, i.e., a higher energy flux towards the colder reservoir. On the other hand, if $\ko>0.4$ the growth of $\Jr$ diminishes, whereas $\Jl$ keeps increasing its magnitude.

\begin{figure}\centering
\includegraphics[width=0.85\linewidth,angle=0.0]{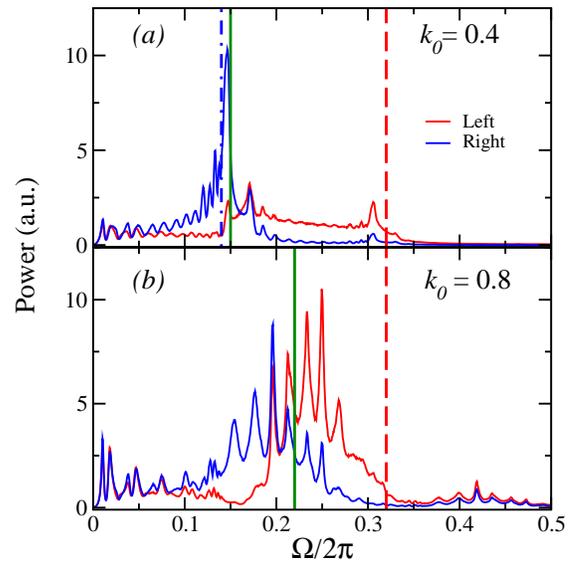}
\caption{(Color online) (a) Power spectra of the two oscillators in the left (red) and right (blue) sides of the contact for a value of the interface elastic constant $\ko=0.4$. (b) Same as (b) but for $\ko=0.8$. Vertical dashed and dot-dashed lines correspond to the cut-off frequencies of the left and right phonon bands, respectively. Vertical solid line denotes the corresponding $\Om/2\pi$ values in each case. Same $\Vl$, $\kl$, $\lambda$, $N$, and $\nc$ values as in Fig.~\ref{fig:2}.}
\label{fig:10}
\end{figure}

The change in relative magnitude of $\Jl$ and $\Jr$ as $\ko$ increases can be explained by examining the power spectra corresponding to low and high values of the interface elastic constant that are plotted in Fig.~\ref{fig:10}; in both instances the left and right phonon bands are $0<\Omega/2\pi\lesssim0.32$ and $0<\Omega/2\pi\lesssim0.14$, respectively. For $\ko=0.4$ more power is available in the right spectrum at high frequencies within the region wherein the phonon bands overlap. Thus the external driving interacts more closely with the right segment, resulting in a well defined thermal resonance with $\Om/2\pi\sim0.15$ and, as a consequence, $\Jr>\Jl$. Next, by examining the spectra for the $\ko=0.8$ case depicted in Fig.~\ref{fig:10}(b) it can be observed that they become entangled and begin to form a whole. This behavior can be explained by noticing that, for this high $\ko$ value, both segments interact more strongly and act as a single system. Therefore a single phonon band, i.e. $0<\Omega/2\pi\lesssim0.32$, determines the frequency values of the noninteracting phonons within the two segments. Thus the ensuing resonant frequency $\Om/2\pi\sim0.22$ lies within the aforementioned phonon band and in a frequency range wherein the overlap of the two spectra is largest. Now, as these can still be distinguished from one another, it is clear that more power is provided by the left spectrum. This in turn entails $\Jl>\Jr$, in agreement with the results of Fig.~\ref{fig:5}.

Finally, we would like to discuss the possibility of an experimental realization of this device. For typical atom lattices, room temperature, measured in kelvins, corresponds to a dimensionless temperature $T_{_0}\in[0.1,1]$~\cite{Li12}; thus the employed $T_{_0}$ value is within the adequate range considering future technological applications. For a lattice constant of 1 \r{A}, a lattice of the size herein used is about 32 nanometers long, a size scale within the reach of current technology in order to be built. Furthermore, at molecular levels a modulation of the coupling between two molecules can be achieved experimentally in molecular junctions by, for example, harmonically varying the distance among them, therefore modulating the coupling between the molecules.

\section{Concluding remarks\label{sec:Disc}}

To summarize: we have studied energy transport control in a one-dimensional segmented system composed of two FK lattices connected by a time modulated coupling. This model affords a convenient way to study dynamical control of heat transport and obtain results that might be relevant for nanoscale devices. Our analysis reveals that, as far as the resonance property is concerned, there is much similarity between the FK and the harmonic models and is consistent with previous findings on the subject~\cite{Zhang11}. By an appropriate scaling we have reduced the number of parameters involved in the description of the system and determined that the phonon heat transport properties in the harmonic limit of the undriven lattice are crucial to control the heat flux through the segments when thermal resonance is present. If values of the amplitude of the on-site potential and harmonic constant in the left segment are greater than those in the right one, energy flow into the colder reservoir is greater than the flow into the hot one, whereas the reverse is true in the opposite case. Also, heat flow into the colder reservoir is higher than into the hotter one for low values of the harmonic coupling constant between segments; the opposite being true for higher values of that same constant. We expect that our results hold for larger system sizes since the herein studied system is essentially the same as that in Ref.~\cite{Beraha15}. Therefore, the shift to lower frequencies of the resonant frequency value reported in the latter work can be expected as the system size increases.

Previously the thermal transport properties of a harmonic lattice system consisting of two semi-infinite leads at different temperature and connected by a time-modulated coupling were studied by means of the nonequilibrium Green's function formalism~\cite{Cuansing10}. There is a net energy flow out of the warmer lead, but for the colder lead energy flow direction depends on the values of both the driving frequency and temperature. Later, a model similar to that herein employed and in Ref.~\cite{Beraha15}, but with the thermal reservoirs modeled as infinite harmonic lattices and the on-site potential being harmonic as well~\cite{Beraha16}, was proposed and studied analytically by the same technique as in the last reference. One of the transport regimes studied corresponds to a heat pump against the imposed thermal gradient. It would be interesting to explore the possibility of obtaining the same effect with the anharmonic model herein employed. 

\smallskip
\begin{acknowledgments}
The author thanks CONACYT, M\'exico for financial support, Brandon Armando Mat\'\i nez Torres for his help in rendering the figures, and Juan M.~L\'opez and Susana Sanchez Merodio for insightful comments and discussions.
\end{acknowledgments}


\bibliographystyle{prsty}

\end{document}